\documentclass[twocolumn]{revtex4}

\usepackage{graphicx}
\usepackage{epigraph}
\usepackage{simplemargins}
\usepackage{float}

\setleftmargin{1.5cm}
\setrightmargin{1.5cm}

\begin{document}
\title{The chaos within: exploring noise in cellular biology}
\author{Iain G. Johnston}
\affiliation{Systems \& Signals Group, University of Oxford, \\Clarendon Laboratory, Parks Road, Oxford, UK}

\maketitle

\epigraph{I say unto you: a man must have chaos yet within him to be able to give birth to a dancing star. I say unto you: ye have chaos yet within you...}{Nietzsche, Thus Spake Zarathustra}

\section*{A brewery in a bouncy castle}
Cellular biology exists embedded in a world dominated
by random dynamics and chance. Many vital molecules
and pieces of cellular machinery diffuse within cells,
moving along random trajectories as they collide with
the other biomolecular inhabitants of the cell. Cellular components may block each other's progress, be
produced or degraded at random times, and become
unevenly separated as cells grow and divide. Cellular
behaviour, including important features of stem cells,
tumours and infectious bacteria, is profoundly influenced by the chaos which is the environment within
the cell walls.

How can the delicate processes that give rise to life
take place in this random world? And what can statistics tell us about the probability of things going wrong?
The study of \emph{cellular noise} -- the causes and effects of
randomness within cellular biology -- is a rapidly growing area within biostatistics attempting to describe these
phenomena.

Modern statistical methods and the explosion of
recent results from experimental biology are allowing
us to understand this essential randomness of cellular
systems in hitherto unrivalled detail. Here we will look
at some important causes and effects of randomness in
cellular biology, and some ways in which researchers,
helped by the vast amounts of data that are now flowing
in, have made progress in describing the randomness of
nature.

\section*{Cities built on shaky ground}
The inner workings of a cell can roughly be pictured as an industrial city, with many different
processes contributing to the city's well-being. Among
these are `power stations' which produce fuel that other
industries harness, a central library where the blueprints
for useful machinery are stored, and factories which produce these machines. Cellular machines -- we call them
proteins -- perform many of the tasks we view as essential
to life: the digestion of food (machines chemically break
down nutrients); movement (machines in muscle fibres
exert forces on each other to move that fibre); production
of energy (machines that create chemical fuels) and so on. In an ideal world, the city would produce
copies of the information in the library and
distribute them to factories, which would read
them and produce these essential machines as
needed, enabling the city to function.
In this metaphor the central library is our
DNA, power stations are our mitochondria,
and factories our ribosomes. The machines, as
we have said, are proteins, the library's books
are genes (each containing the instructions on
how to build a protein) and the copies of those
books are mRNA molecules, which convey
this information and are `translated' to produce
proteins. This process, illustrated in Fig \ref{fig1}a,
is often referred to as the \emph{central dogma of
cellular biology}: genes are first transcribed to
mRNA, then translated to form proteins, the
building blocks of the cell.

However, our cities are very unpredictable places. First of all, things fall apart rather
quickly. The copies of building instructions --
the mRNAs -- are particularly prone to this.
Worse, the library only makes some of its
books accessible at a time. The unpredictable
opening and closing of books, and random
nature of production and degradation in our
metaphorical cities, are inevitable consequences of the random dynamics in cells: in biology,
these processes all involve chance collisions
and rearrangements of molecules within the
chaotic interiors of cells.
So, if we happen to find a book open
and make a copy of its contents, we can make
several of the corresponding protein machines
-- but the book may close and the copies may
degrade very soon, and we are stuck with this
small number. (This is the \emph{copy number} of the
protein, and it can range between dozens in a
cell and thousands.) Our city may require a
particular machine -- a particular protein -- with
some urgency, but if the corresponding book
only opens rarely and the instruction copies
degrade quickly, we may be unable to produce
that protein in sufficient numbers to function.
If books open and close several times, we will
see unpredictable `bursts' of production in the
cell. The copy number of a cellular machine,
dependent on these random processes, is therefore uncertain, giving rise to a spread of possible
values at any time, and leading to variability in a
city's ability to perform biological tasks.

\begin{figure}
\includegraphics[width=8cm]{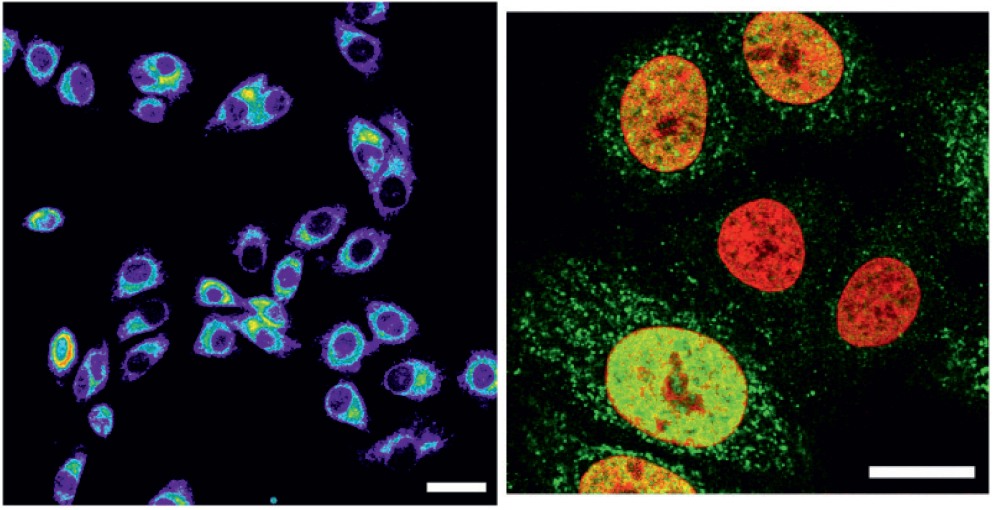}
\caption{\footnotesize{\textbf{Picturing noise in cellular biology.} Left: Measurements of the mitochondrial content of cells (yellow is high
mitochondrial density, purple is low) showing significant variability in the number of `power
stations' between otherwise similar cells. Right: Green speckles are sites where transcription is
taking place (the first stage of gene expression). Some cells -- like the large yellow one -- transcribe
quickly, producing more cellular machinery, whereas some -- like the redder ones -- show very little
transcriptional activity. Images from Ref. \cite{dasneves}.}}
\label{fig0}
\end{figure}

\begin{figure}
\includegraphics[width=10cm]{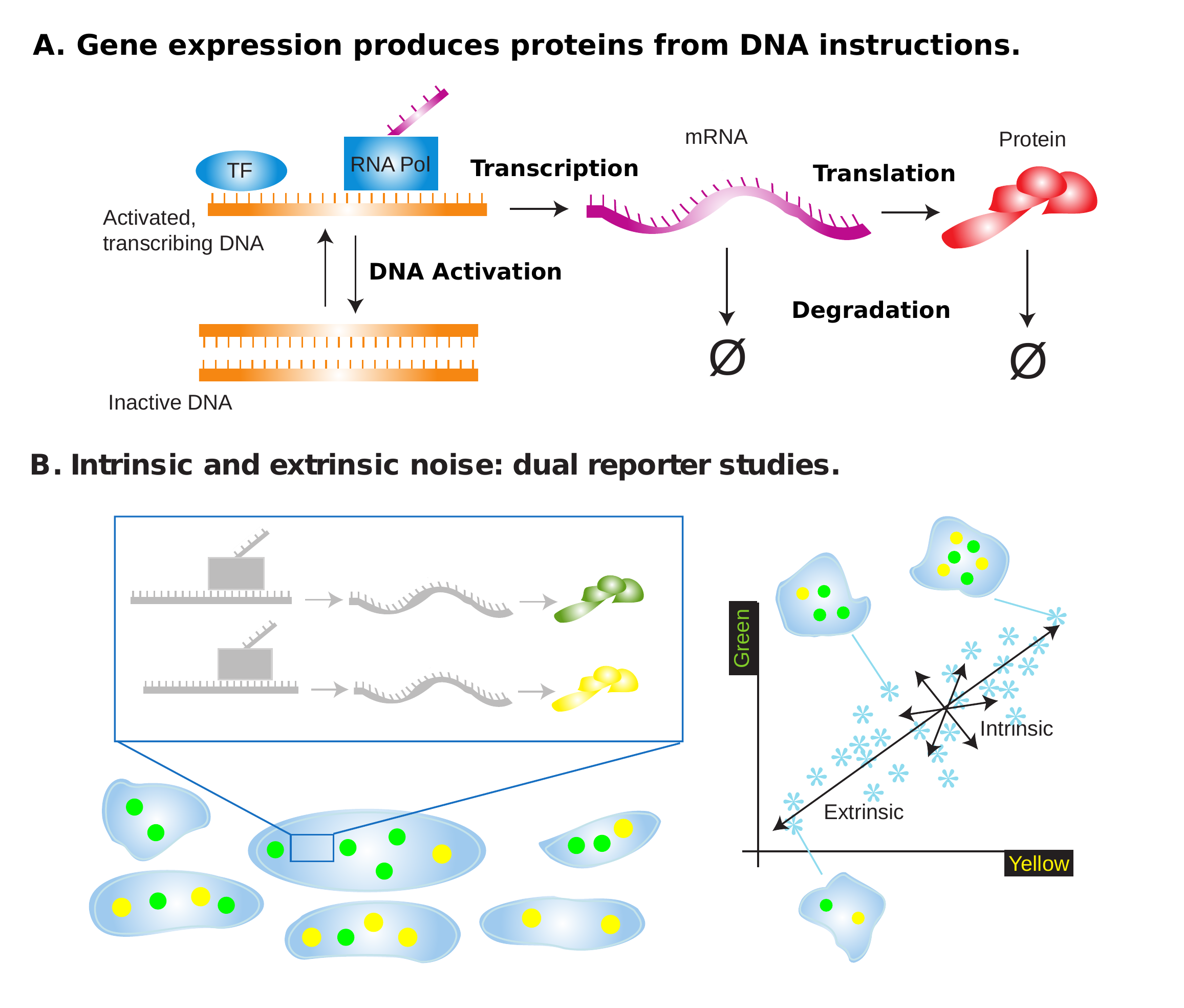}
\caption{\footnotesize{\textbf{Noise in gene expression.} A. The `central dogma' of cell biology, whereby DNA is activated and transcribed to produce mRNA, which is translated to produce proteins. In the metaphor in the main text, this process is represented by the opening of library books which are interpreted to produce blueprints, which are distributed and assembled to produce machinery. B. Quantifying intrinsic and extrinsic noise in a population of cells. In the graph the long diagonal line shows variations in overall brightness, due to extrinsic noise;
cells near its origin have few fluorescing genes, those near the end have many. The shorter arrows show cells
glowing more green or more yellow; their variation is not in number of genes, but in the proportion of green
and yellow ones.}}
\label{fig1}
\end{figure}

\section*{Genes rolling dice}
Much of the existing work on cellular noise
has considered variability in \emph{gene expression} -- the levels at which the products of genes are
present within cells. Most of the cells in our
bodies are genetically identical: for example,
muscle cells are genetically identical to brain
cells. But the two are very different in appearance, behaviour and biochemical profile. A
fundamental reason for this is the differences
in gene expression within different cells: although cells may contain the same genetic information, only a subset of genes are expressed -- `turned on' -- in any given cell, and the ones
that are turned on determine what the cell is
and what it does. The genes expressed in a cell
are an important determining factor of its `cell
type' (including its appearance, behaviour, and
other attributes), allowing different cells to
fulfill different roles in our bodies.
To express this in terms of our metaphor,
some city libraries may intentionally keep books
on particular machines open more than others,
so that, for example, one cell-city produces lots of
proteins that process raw materials, and another
produces proteins that facilitate movement.

However, even in cells of the same type
(with the same library patterns), cell-to-cell
differences in gene expression still occur, provoked by random differences in features like
cellular size, available energy levels and chemical environments. These cell-to-cell differences
exist alongside the within-cell differences in
gene expression that we met previously.
We normally refer to random differences
within a cell as \emph{intrinsic noise} and cell-to-cell
(city-to-city) differences as \emph{extrinsic noise}.
Noise in this context has a specific statistical meaning: it is most often defined as the
\emph{coefficient of variation} of a quantity, the standard deviation of a distribution divided by its
mean. Typical noise levels in gene expression
levels can be as high as 0.4 -- the standard deviation in is nearly half the mean value (Table \ref{table1}).

\begin{figure}
\includegraphics[width=8cm]{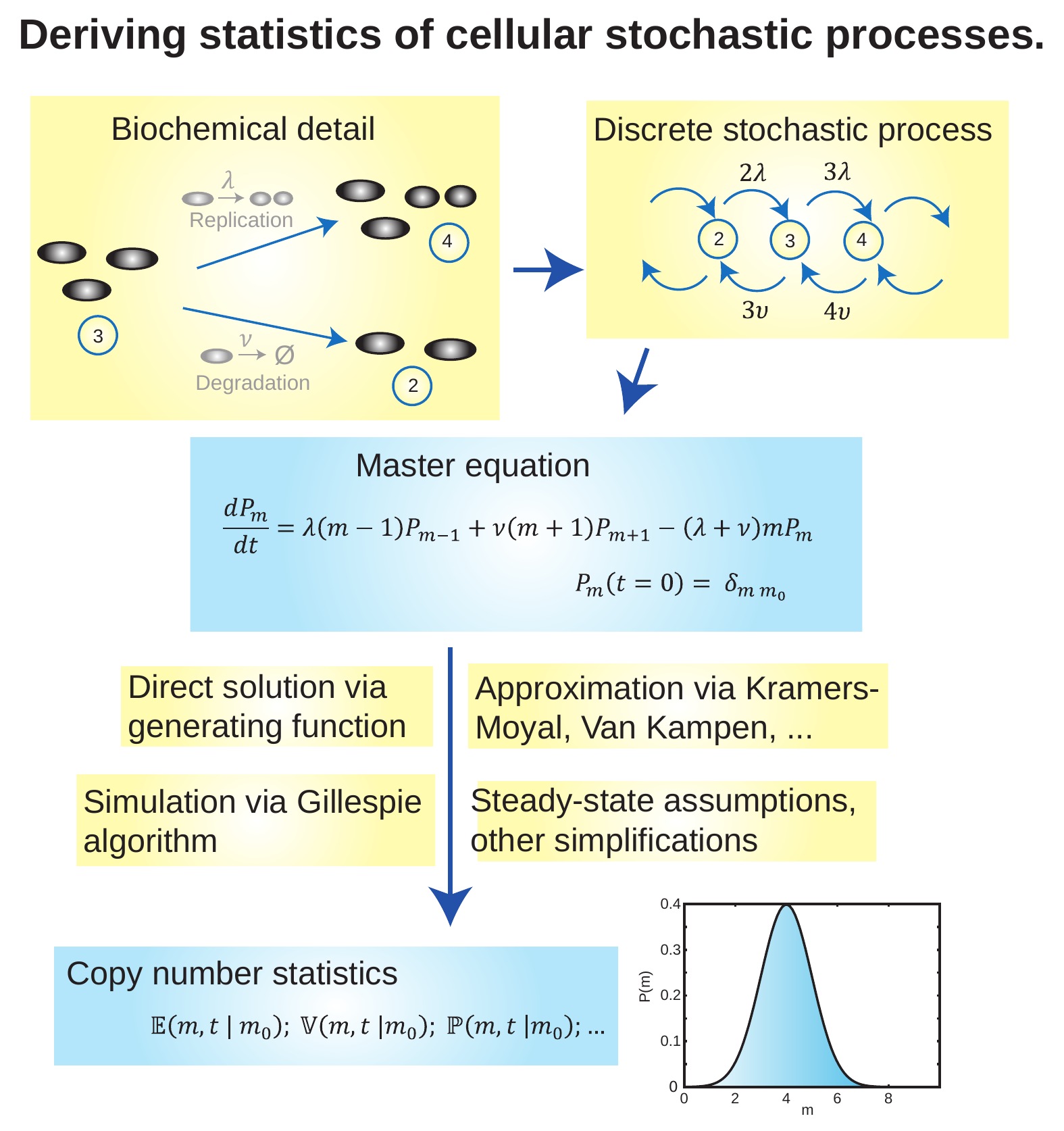}
\caption{\footnotesize{\textbf{Deriving statistics of cellular stochastic processes.} A stochastic model of a biochemical process is created by considering the events that can change the biochemical state of a system. This model is written down as a master equation, describing the time evolution of the probability with which the system will be in a given state. Analytic and computational tools are then used to derive statistics for bionumbers of interest from this central equation.}}
\label{fig2}
\end{figure}

Biologists have explored these types of cellular noise
in elegant experiments -- the first of which, in
2002, kicked off interest in cellular noise \cite{elowitz}.
Picture two genes X and Y under identical
regulation in a cell, so that in a perfectly deterministic environment we would expect equal
levels of X and Y in each cell in a population.
Elowitz \emph{et al.} inserted two such genes into \emph{E.
coli} cells -- the genes were identical except that X
glowed yellow and Y glowed green. They found
that some cells glowed with similar brightnesses
but in different colours -- some more green
and some more yellow, with cells producing,
randomly, more copies of one or of the other (see Fig. \ref{fig2}b).
This is intrinsic noise: there were different proportions of X and Y within each cell. But some
cells were overall very dark or very bright: some
cells (perhaps with more energy) were making
more copies of both genes, so that there were different total amounts of X and Y between cells.
This is extrinsic noise. These measurements
of noise levels in gene expression for the first
time showed how pronounced the variability
is in this fundamentally important biological
process. Elowitz \emph{et al.} were able to quantify the
noise, and thus to look at the statistics of protein
production in a population of cells.

\begin{table*}
\footnotesize
\begin{tabular}{l|l|l|l|l}
& Intrinsic & Source (organism) & Extrinsic & Source (organism) \\
\hline
Prokaryotic genes & 0.2 & Elowitz \cite{elowitz} (\emph{E. coli}) & 0.3 & Elowitz \cite{elowitz} (\emph{E. coli}) \\
Eukaryotic genes & 0.05-0.2 & Newman \cite{newman} (budding yeast) & 0.1-0.4 & Newman \cite{newman} (budding yeast) \\
& 0.01-0.05 & Raser \cite{raser} (budding yeast) & 0.1 & Raser \cite{raser} (budding yeast) \\
Cell volume & N/A & & 0.07 & Volfson \cite{volfson} (theoretical) \\
Mitochondrial mass & N/A & & 0.32 & das Neves \cite{dasneves} (HeLa) \\
Mitochondrial membrane potential & 0.2-0.3 & Collins \cite{collins} (HeLa) & 0.25 & das Neves \cite{dasneves} (HeLa) \\
Transcription rate & ? & (no studies yet) & 0.4 & das Neves \cite{dasneves} (HeLa) 
\end{tabular}
\normalsize
\caption{\footnotesize{\textbf{Magnitudes of cellular noise.} Approximate ratios of the standard deviation to the mean for several biological distributions.}}
\label{table1}
\end{table*}

\section*{Other players}
It is not just levels of gene expression that
differ between cells. The concentration of
chemicals, such as sources of nutrition or
oxygen, may vary across a cell or a population
of cells, causing extrinsic differences. If we
have a population of dividing cells that are not
synchronised, we would expect extrinsic differences in cell size (as cells grow and divide).
Partitioning noise, whereby the two daughters
of a parent cell inherit different amounts of
component proteins and organelles, also leads
to extrinsic differences in a population. The
physical environment that a cell occupies is another potential source of extrinsic variability:
cells in the centre of a connected population,
for example, may be under higher physical
pressure from their confinement than cells on
the edge of a colony; those on the edge may
be surrounded by more or by less nutrient. All
these factors may have important downstream
effects on cellular behaviour.

\section*{Capricious, chaotic cells}
As well as being the fundamental active
elements of cellular machinery, proteins are
responsible for transmitting many signals
within the cell. These signals may affect the
production of other proteins (opening or
closing books in the library), so noise in the
production of one type of protein can affect
the production of many others. Even small
random effects can amplify to produce radical
cell-wide effects.

An important medical example concerns
the action of a drug called TRAIL, which kills
cancer cells by starting a chain of messagesending within cells: one protein activates
another protein which activates another, with
the end result being the triggering of processes
which kill the cell. However, as proteins are
the medium through which these messages are
passed, differences in protein levels between
cells create differences in the strength of the
message, leading to some tumour cells being
killed quickly and some persisting for much
longer \cite{spencer}. This is an example of \emph{fractional killing},
whereby each round of treatment kills some
but not all of the cells within a tumour; it is of
great importance in cancer therapy, and extrinsic noise is being increasingly implicated as a
source of this statistical variability.

\section*{No perfect cell}
Important cellular control processes are also
performed using signals transmitted by proteins. Proteins are subject to fluctuations and
random effects, so no cellular process
can ever be controlled perfectly. The fundamental limits that biological noise sets on a
cell's ability to control its biochemical contents
were recently described in a fascinating merger
of information theory and biostatistics \cite{lestas}. This work essentially constitutes a fundamental law of biological information processing, proving a lower
limit on the error emerging from biological
control processes.

How do cells deal with these fundamental limitations on their ability to control
what they do? Many regulatory mechanisms
within cells have evolved architectures designed to reduce noise or allow a limited degree of
control \cite{maheshri}. Negative feedback loops abound in
cellular circuitry, allowing fluctuations to be
damped and perturbations to be reduced.
Some mechanisms have even evolved to take
advantage of noise. In `bet-hedging' in bacterial populations, genetic `switches' within
bacteria respond to random cues, so that
some members of a population are switched
into an active, infectious phase and others
into a robust, quiescent phase \cite{fraser}. Antibiotic
treatments may kill many of the active bacteria, but the robust quiescent subset of the
bacteria survives for longer, allowing the
infection to weather the storm and propagate
in the future.

These examples are the tip of the iceberg
of the effects of cellular noise. The production
of different tissue types and the energy levels
within cells are all subject to random influences, as are a host of other processes, with
more being elucidated every day.

\onecolumngrid
\begin{figure}[H]
\includegraphics[width=18cm]{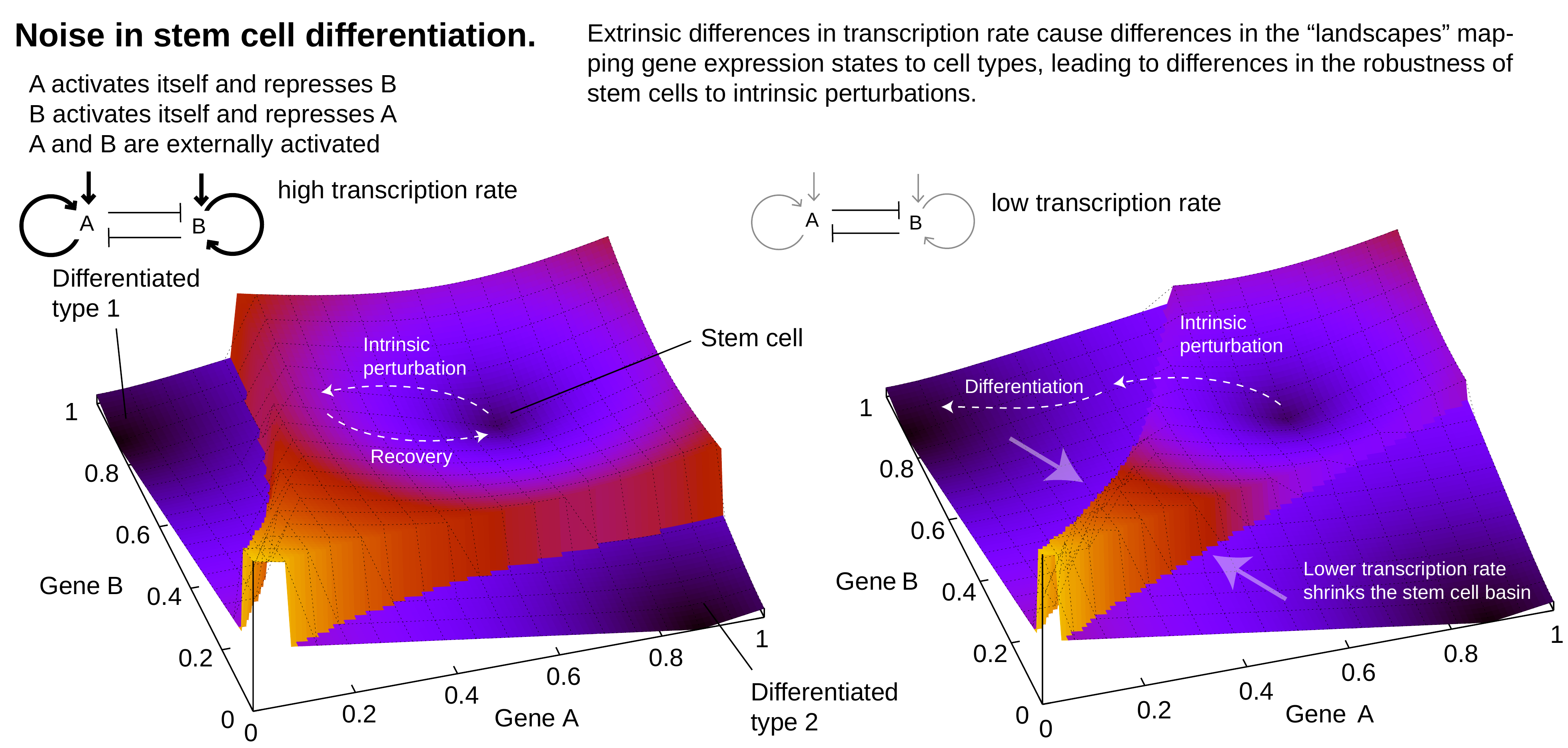}
\caption{\footnotesize{\textbf{Noise in stem cell differentiation.} These `landscapes' relate the expression levels of gene A and B to different cell states. The stable region associated with the `stem cell' state is wider for cells with high transcription rate, so this state is less sensitive to intrinsic noise in gene expression.}}
\label{fig3}
\end{figure}
\twocolumngrid 

\section*{Master equations and random walks}
The importance of randomness in gene expression has led to a mathematical model that
is rapidly becoming canonical \cite{paulsson}. The binding,
unbinding, production and degradation processes within a cell are modelled as Poisson
processes: events occur randomly and independently, with certain average rates, which are parameters of the model. Similar models can
be constructed for features other than genes,
representing the production and degradation
of cellular constituents as random processes
with rates to be determined.
From a mathematical model of a cell's
random behaviour we can calculate, using
standard techniques, important bio-numbers
-- statistics like the expected copy number
of cellular components over time, and the
variability we may expect in this value between
cells. A \emph{master equation} can be written down,
describing the probabilities of observing different states of the system (for example, a cell
containing 50 mRNAs) and how these change
with time. For the mathematically inclined, a
schematic of the derivation of this equation is
shown in Fig. \ref{fig2}.

If a system is so complicated
that analytic progress is impossible we can construct numerical models within a computer. Many incarnations of any randomly reacting
system can be realised in this way, explored
numerically, and the statistics of the resulting
ensemble can be found. Using recent advances
in the statistical field of \emph{parametric inference}, we
can make a `probability landscape' describing
possible values for cellular bio-numbers and,
importantly, suggest experimental designs
that will help tighten these probability distributions and get a better handle on the realworld parameter values. In this way, as well
as discovering important numbers in cellular
biology, statistics can inform experimental
biology about the most valuable experimental
approaches, where the smallest effort can be
used to get the greatest reward.

\section*{Fluctuating power stations}
Our own work focuses on mitochondria -- our
cities' power stations -- as an important source
of cellular noise. Since mitochondria provide
ATP, a fundamental energy source for cells,
variability in their presence or performance
can have dramatic effects on a wide range of
cellular phenomena.

Mitochondria grow and divide, are removed by the cell if they perform poorly, and
are inherited (in randomly different proportions) as cells themselves divide. The dynamics
by which mitochondria are inherited and by
which they grow and propagate naturally leads
to variability in the size and functionality of
mitochondrial populations within cells \cite{johnston}. Cells
with few, or poorly-functioning
mitochondria, have lower levels of ATP and
their internal processes (including protein
production) are slower as a result. This extrinsic variability has been experimentally linked
to differences in transcription rates between
cells, and, through its effect on energy levels,
is theoretically predicted to affect a host of
downstream phenomena.

\section*{Noise in stem cell differentiation}

An example of an important predicted
consequence of mitochondrial variability
concerns stem cell differentiation \cite{johnston}. Stem cells
are cells that can divide and produce other cell
types: for example, a blood stem cell may after
several divisions produce a red or a white blood
cell as well as many other alternatives. The cellular decision to produce a particular cell type
is made through expression levels of certain
characteristic genes: for example, high expression of gene A and low expression of gene B
may correspond to a red blood cell, low A and
high B to a white blood cell, and intermediate
levels of both to an undifferentiated stem cell.
(And, as we have seen in the yellow and green
fluorescent example above, these gene expression profiles are subject to a degree of chance.)
These relationships give rise to a `landscape' mapping gene expression levels to cell
states. These landscapes are often thought of
in terms of basins -- regions, like the drainage
basin of a river, where all paths flow downhill
towards a final stable state. Some gene expression profiles are more stable than others -- a stem cell, for example, that experiences a small
perturbation in expression of gene A from
intrinsic noise will `flow downhill' back into its
own basin and recover its original expression
profile without being forced into a different
state -- it will remain a stem cell. This stability
arises from negative feedback as mentioned
above: an example of the robustness of cellular
biology to intrinsic noise.
However, extrinsic variability in transcription can change the structure of the landscape,
making different cell states more stable or less.
If transcription rate is decreased (perhaps due
to a lower mitochondrial content) in Fig.
\ref{fig3}, for example, the basin containing the stem
cell state shrinks, and a perturbation is more
likely to knock the system into an adjacent
basin, from which it will flow downhill into a
new stable state, corresponding to a differentiated cell type -- our stem cell may become a
red blood cell. Consequently, the properties
of the mitochondrial populations within stem
cells affect whether making either stem cells or
differentiated daughters is more likely \cite{johnston} -- an
important factor during the development of
organisms and in the correct maintenance of
cell populations throughout life.

\section*{Taming the chaos within}

Cellular biology is embedded within a noisy,
chaotic world, where random effects influence many vital processes, with important
consequences in fields extending from fundamental
biology to medicine. The recent development of
experimental, analytic and computational tools
to explore noise in cellular biology is, for the
first time, allowing us to explore the causes and
effects of these random influences. Physicists,
biologists, mathematicians and statisticians are
working together to probe the random nature
of cellular biology and create a consistent way of
finding the probabilities, time scales and mechanisms associated with nature's rolls of the dice.
For further reading on this expanding
field, many excellent review articles exist on
cellular noise -- see below and references therein.

\bibliographystyle{unsrt}
\bibliography{refs}

\end{document}